\documentstyle[12pt, epsf]{article}
\newcommand{\be}{\begin{equation}}
\newcommand{\ee}{\end{equation}}
\newcommand{\ba}{\begin{eqnarray}}
\newcommand{\ea}{\end{eqnarray}}

\newcommand{\hmp}{${\rm Mpc}$/h}

\newcommand{\etal}{et al.~}
\newcommand{\eg}{e.g.,~}
\newcommand{\ie}{i.e.,~}
\newcommand{\lcdm}{$\Lambda$CDM}

\begin{document}
\title{Tessellating the Universe: the Zel'dovich and Adhesion tiling of space}
\author{Sergei F. Shandarin\\
Department of Physics and Astronomy\\
University of Kansas\\
Lawrence, KS, USA}
\maketitle
\begin{abstract}
The adhesion approximation is a simple analytical model suggested for explanation of
the major geometrical features of the observed structure in the galaxy distribution 
on scales from 1 to (a few)$\times$100 \hmp. It is based on Burgers'
equation and therefore allows analysis in considerable detail. A particular version
of the model that assumes the infinitesimal viscosity naturally results in irregular tessellation 
of the universe. Generic elements of the tessellation: vertices, edges, faces and three-dimensional
tiles can be
associated with astronomical objects of different kinds: clusters, superclusters and voids of galaxies.
Point-like vertices   contain the most of the mass and 
one-dimensional edges (filaments) are the second massive elements. The least massive are 
the two-dimensional faces and tiles (voids).  
The evolution of the large-scale structure can be viewed  
as a continuous process  that transports mass predominantly from the high- to low-dimensional elements of the tessellation.  For instance, the mass from the cells flows into faces, edges and
 vertices, in turn the mass from faces flows into edges and vertices, etc.  
 At the same time, the elements of the tessellation themselves are in continuous motion
 resulting in mergers of some vertices, growth of some tiles and shrinking and disappearance
  of the others as well as other metamorphoses.
\end{abstract}
\section{Introduction}
We live in an expanding universe theoretically predicted by Russian mathematician 
A. Friedman in 1922 \cite{fri-22}  and independently  discovered by American astronomer 
E. Hubble in 1929  \cite{hub-29}. It means that in  the perfectly homogeneous universe 
 the relative velocity, ${ v}$, between any two particles or objects would be described by 
 the Hubble law
\be
{v} = H(t) {r},
\label{eq:hubble-law}
\ee
where ${r}$  is the distance between  the objects, the velocity is directed along the straight line
connecting the objects. In the real inhomogeneous universe
this law can be applied only at sufficiently large distances. A positive function $H(t)$ is called the Hubble parameter, it characterizes the rate of expansion of the universe. Its present value 
$H_0 =H(t_0) \approx$ 71 km/s/Mpc (Mpc stands for megaparsec, 1 Mpc = 10$^6$ pc $\approx$
3,260,000 light years). Until the end of the millennium most cosmologists 
believed that the expansion of the universe was faster in the past and has been
monotonically slowing down with time (\ie $\dot{H} < 0$ at $t>0$)  
due to the pulling effect of gravity. However the study of extremely distant supernovae 
stars revealed that the 
expansion of the universe recently (by cosmological standards) started to accelerate 
\cite{sn}. 

The acceleration of the universe can be easily explained if the most of mass in the universe is in the
form of dark energy -  a hypothetical form of energy uniformly filling all the space and experiencing
a repulsive gravitational force. Although the physical nature of dark energy remains a mystery it
is currently the most popular way of explaining the acceleration of the expansion of the
universe. According to current measurements of the cosmological parameters dark energy
accounts for 74\% of the total mass-energy, 22\% are due to dark matter - another puzzling 
component, while the ordinary matter often dubbed as baryonic matter makes only about 4\% of
the total mass. Dark matter is believed to consist of weakly interacting massive particles that
interact with ordinary matter and themselves via gravity only  for most intents and purposes.
Although the exact kind of particles has been neither identified nor detected 
in physical experiments the hypothesis of dark matter is well founded.
The gravitational interaction of dark matter is similar to that of ordinary matter. Dark matter plays
a crucial role in the formation of the structure in the universe due to its dominance in density
and therefore in generating attractive gravitational force. The universe is also permeated by the Cosmic Microwave Background (CMB)
radiation that brings unique information about tiny initial perturbations of the uniformity of the 
universe that served as the seeds for the presently observed structure in the universe. However
CMB radiation makes less than 0.01\% of the total mass of the universe at present and has played 
a very limited role in the structure formation after the universe was roughly one thousandth
of its present size. It has been also established that  the geometry of the universe is very close to flat.
\begin{figure}[h]
    \epsfxsize = 6 truein
    \hskip .65 truein
    \epsfbox{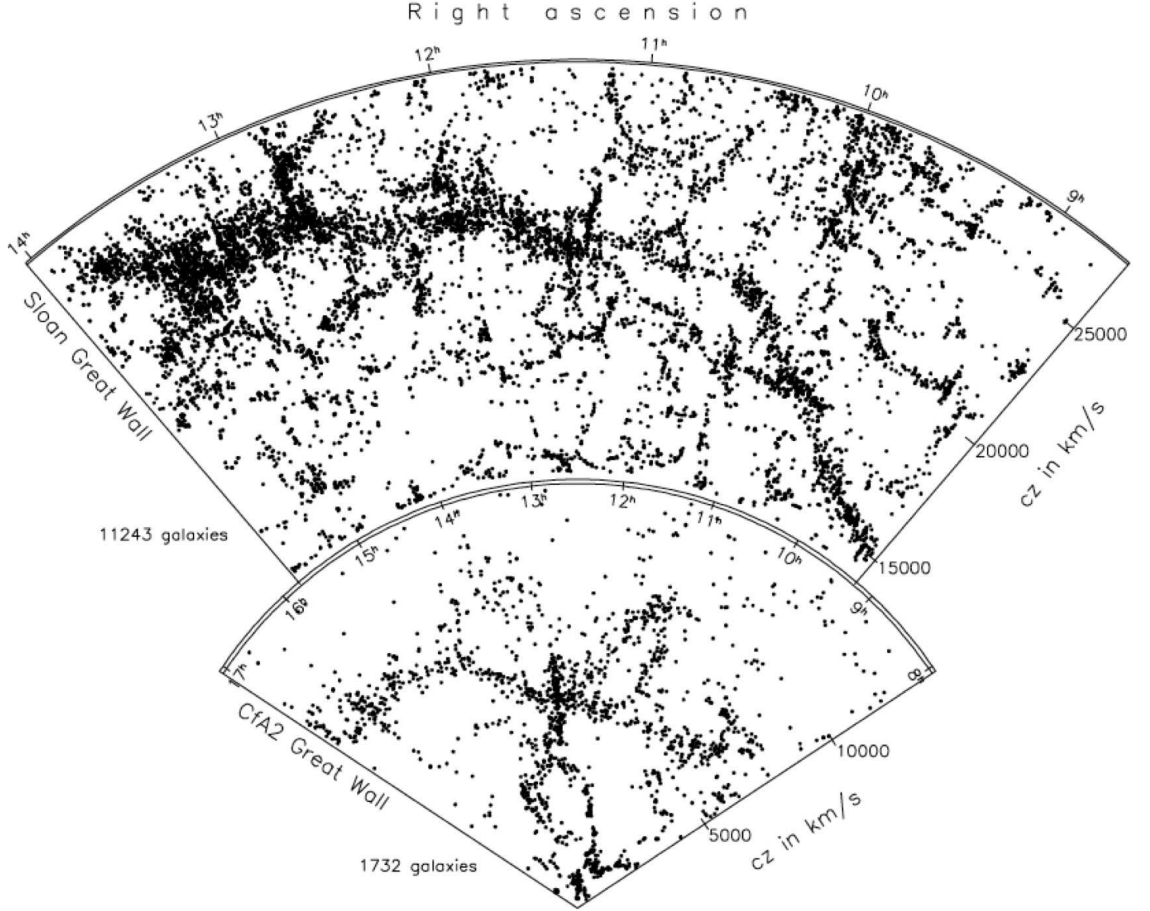}
    \caption{\small Sloan Digital Sky Survey Great Wall (top) compared to CfA2 Great Wall (bottom) 
    at the same scale. The scale is shown in the velocity units (km/s), the distance in
    Mpc can be obtained from the Hubble law (Eq. \ref{eq:hubble-law}), \eg 10,000 km/s correspond
    to 141 Mpc. Each point represents a galaxy. Adopted from \cite{got-etal-05}.} 
    \label{fig:GWs}
\end{figure}
The present universe is highly inhomogeneous on scales up to several hundred of Mpc 
as shown in Fig. \ref{fig:GWs}. 
However, this figure does not show a true distribution of galaxies in physical space because
the line of sight distances have been derived from the receding velocities via the Hubble
law (Eq. \ref{eq:hubble-law}).  In the inhomogeneous universe the actual relation between
the line of sight distance, $r$, and recession velocity, $v_{obs}$, is given  by the equation
\be
{v_{obs}} = H_0 {r} + v_p,
\label{eq:vel-dist}
\ee
that involves an additional term, $v_p$, called the peculiar velocity. The peculiar velocities are
 due to growth of density  inhomogeneities in the universe that obviously requires a transport 
 of  mass from one place to  another and therefore demands additional 
 (with respect to uniform expansion) velocities. 
The recession velocities of galaxies, $v_{obs}$, can be measured by making use of
the Doppler effect, while the peculiar velocities cannot be easily measured. Therefore,
the line of sight distance derived from the observed velocity $r_{est} = v_{obs}/H_0 = r + v_p/H_0$ 
gives only a crude estimate of the  true distances of galaxies.

\begin{figure}[h]
    \epsfxsize = 6 truein
    \hskip .65 truein
    \epsfbox{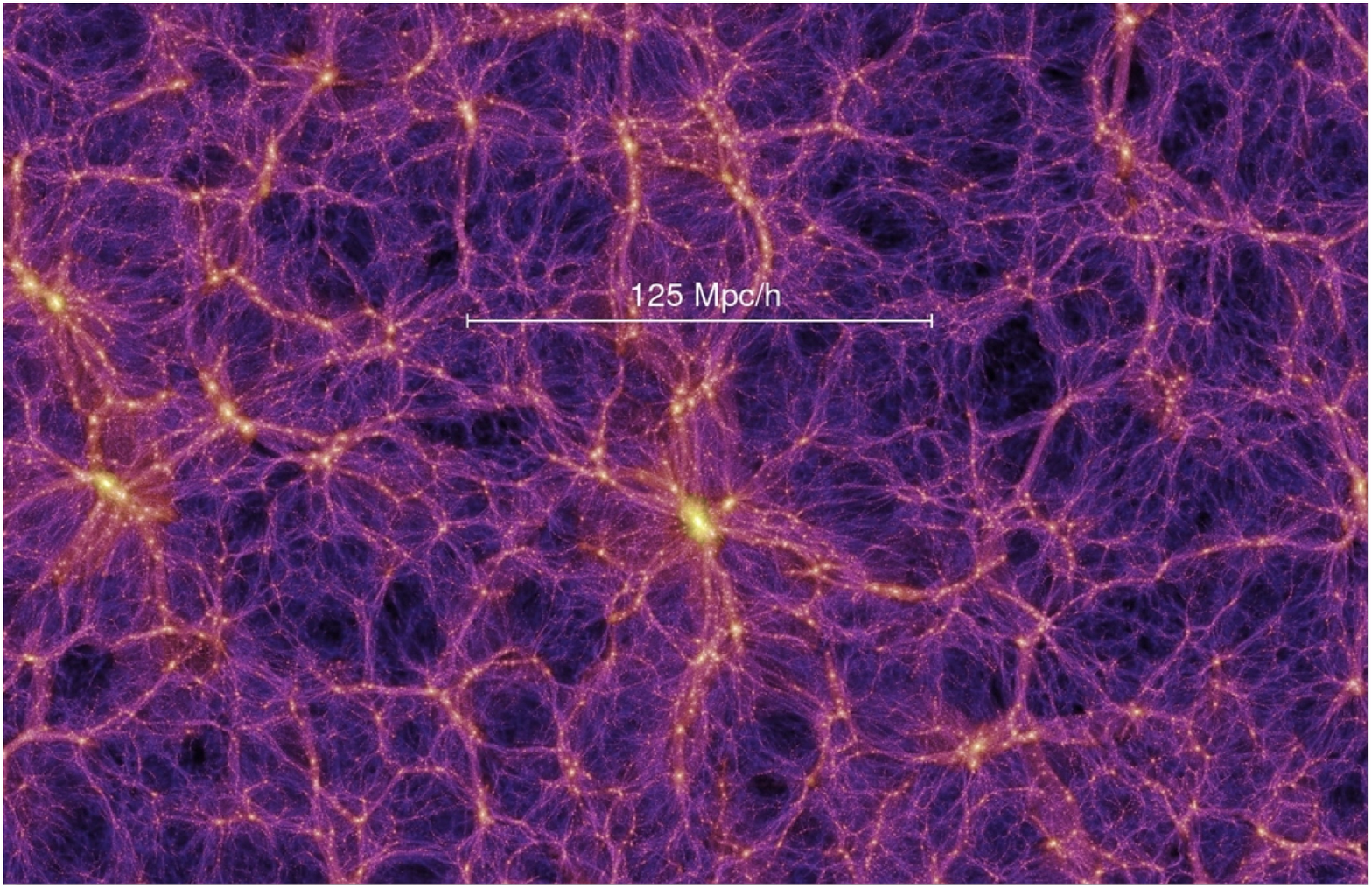}
    \caption{\small The dark matter density field in a slab of thickness 15 \hmp \, obtained in
    the Millennium simulation \cite{virgo98} of the structure formation in the universe.} 
    \label{fig:millennium}
\end{figure}

The density distribution in physical space is better illustrated (albeit only in statistical sense) 
by the N-body simulations of the structure formation in the currently popular cosmological 
model referred to as the  Lambda Cold Dark Matter ($\Lambda$CDM) model shown in
Fig. \ref{fig:millennium}.
In cosmology the term large-scale structure is commonly referred to a distribution
of galaxies on the scales roughly from
1 to a few hundred Mpc. Galaxies can not probe smaller 
scales because of discreteness, and on larger scales the galaxy 
distribution is monotonically approaching  homogeneity with the growth of the scale.
The redshift surveys reveal
spectacular abundance of structures often described as filamentary,
network, or bubble structure \cite{gel-huc-89,daCos-etal-94}. 
The  origin  of  the large-scale structure is one of  the  
most important problems  in modern
cosmology. Many fundamental issues in physics, cosmology and
astronomy ranging from speculations on the physical
nature of dark matter, to the measurement of angular anisotropy of the
microwave background radiation as well as establishing the epoch of galaxy 
formation join together here, see \eg \cite{pee-80,sh-z-89}.

Modern theory explains the formation of the structure in the universe
as a result of growth of primeval tiny density fluctuations in the process known as  
gravitational instability. 
Primeval perturbations are assumed to arise as vacuum fluctuations
during the very early stage when the universe was expanding exponentially
(inflationary universe), see \eg \cite{muk-05}. 
Afterwards, the density perturbations had a long
evolution before they become galaxies, clusters of galaxies, 
superclusters and voids of galaxies. The formation of galaxies is a very 
difficult problem on its own. Many complex physical processes like 
star formation and supernova explosions are very important for 
understanding the galaxy formation. We shall
discuss the mass distribution only assuming that galaxies are fairly good 
(though not perfect) tracers of mass on large scales.

As long as the density perturbations are small by amplitude their evolution
is described by the linear theory of gravitational instability 
(see e.g.  \cite{pee-80,sh-z-89}).  
The linear theory is quite simple and therefore is well understood.
In particular, it predicts the rate of growth of the perturbations, $D(t)$: 
$\delta \equiv\delta \rho/ \bar{\rho} \propto
D(t)$, where $ \delta \rho = \rho - \bar{\rho}$ and $<\delta^2>^{1/2}\, \ll 1$.
However,  in commonly used Eulerian form it is not very useful at the nonlinear stage 
when the amplitude of the density perturbations becomes large and the observable structures
(sheets, filaments, and clusters of galaxies) form. 

The analysis of the evolution of perturbations at the nonlinear stage  $<\delta^2>^{1/2}\,\,\ge 1$
becomes quite difficult.
The most straightforward way to address the complexity of nonlinear evolution 
is to carry out three-dimensional N-body simulations (Fig.\ref{fig:millennium}). 
Usually in the simulations of this type the medium
is assumed to consist of collisionless particles, in agreement with
the hypothesis of dark matter in the form of weakly interacting particles. 
The trajectory of each particle in the simulation is numerically integrated 
in the gravitational field generated by all the rest particles. 
Boundary conditions are commonly assumed to be periodic that imitate the infinite universe.

Here we outline another approach to the problem of the large-scale
structure in the universe. We present two approximate analytic solutions 
to the set of PDEs describing the growth of density inhomogeneities in an expanding
universe. One suggested by Zel'dovich in 1970 \cite{z-70} is known as the Zel'dovich approximation
and the other is based on  Burgers' equation \cite{bur-74} and is often referred to as the adhesion 
model \cite{gur-sai-sh-85,gur-sai-sh-89} (see also \cite{sh-z-89}). A particular version of the adhesion
model naturally describes the structure in the universe as an irregular tessellation or quazi-Voronoi
tessellation \cite{mol-etal-97}. The generic 
elements of the tessellation in 3D (vertices, edges, faces, and tiles) can be associated with
the observed structures in the three-dimensional distribution of galaxies (clusters, superclusters and
voids of galaxies).

Both the N-body simulations and approximations require initial conditions. 
In the linear regime the density fluctuations are assumed to be a realization of
a Gaussian random field specified by the power spectrum and the amplitude. 
The current  measurements of the angular fluctuations in the temperature 
of the microwave background radiation by WMAP (Wilkinson Microwave Anisotropy
Probe)  are in excellent agreement with the assumption of Gaussianity. \cite{wmap5}. 
The amplitude of the temperature fluctuations 
suggests that the scale where the density fluctuations have
recently reached nonlinearity is about 3 - 6 Mpc depending on exact
definition of the scale of nonlinearity. This is also in a good
agreement with the observations of the large-scale distribution of galaxies.
The shape of the power spectrum of the initial (\ie linear) density fluctuations
is fully determined by the parameters of the cosmological model. At large scales (small $k$)
it behaves as $P(k) \propto k$ while at very small scales it is approximately $P(k) \propto k^{-3}$
with smooth transition between these limits. It is approximately $P(k) \propto k^{-1.5}$ at the present
scale of nonlinearity, see the discussion bellow and Fig.\ref{fig:power-sp}. 

Here we discuss the models that deal only with the dominating component of the mass
able to cluster \ie the dark matter. The luminous objects in the universe are made of baryons,
therefore it is important to understand the dynamics of the baryonic component of mass
as well. However, this problem is much more complicated and is beyond the scope of this
review. We refer the interested reader to paper \cite{jon-99} that offers
an interesting generalization of the models described here and addresses the 
dynamics of baryons.

The rest of the paper is organized as follows. The equation describing the evolution of density 
inhomogeneities are present in Sec. 2. Sections. 3 and 4 briefly describe the Zel'dovich 
approximation and the adhesion model respectively. Section 5 provides a short summary.

\section{Basic Equations}
The evolution of density inhomogeneities can be described by  a  system
of three partial differential equations 
comprising the continuity, Euler, and the Poisson 
equation (see e.g. \cite{pee-80,sh-z-89}).
In order to exclude the uniform expansion of the universe 
comoving coordinates, {\bf x},  and peculiar velocity, ${\bf u}_p$ are commonly used. 
They are defined by the following relations
\be 
{\bf r}=a(t) {\bf x}, ~~\dot{\bf r} = H(t){\bf r} +  {\bf u}_p.
\ee
A monotonically growing function $a(t)$ called the scale factor 
describes  the uniform expansion of the universe, and $H(t) = d \ln{a}/dt$
is the Hubble parameter introduced by Eq.\ref{eq:hubble-law}. 
The scale factor is completely determined 
by the parameters of the cosmological model.
For instance,  $a(t) \propto t^{2/3}$ in the Einstein - de Sitter universe where 
$\Lambda=0$, and $\Omega=1$, while in more realistic $\Lambda$CDM model  it
is a complicated but known function of time \cite{car-etal-92}. 
In the course of  evolution neither peculiar velocities no gravitational potential
reach the relativistic values, therefore the use of classical mechanics and Newton's gravity law
 is perfectly  appropriate.
In terms of the comoving coordinates and peculiar velocities
the equations describing the gravitational instability in the expanding universe
are as follows: \\
the continuity equation
\begin{equation}
{\partial \rho \over \partial t}  + {1 \over a}
\nabla \cdot ( \rho {\bf u}_p) = - 3H \rho,
\label{eq:cont}
\end{equation}
the Euler equation
\begin{equation}
{\partial{\bf u}_p\over \partial t} 
+ {1 \over a}({\bf u}_p \cdot \nabla){\bf u}_p
= -{1\over a} \nabla \phi - H {\bf u}_p,
\label{eq:euler}
\end{equation}
and the Poisson equation
\begin{equation}
{1 \over a^2} \nabla^2 \phi = 4 \pi G   ( \rho - \overline{\rho} ) ,
\label{eq:poisson}
\end{equation}
where $\rho$ and $\overline{\rho}$ are respectively
the density and mean mass density; $\phi$ is the gravitational
potential due to the inhomogeneities of density; $G$ is the gravitational 
constant. 

The pressure is neglected since we study the medium interacting only
gravitationally. Additional terms on the right hand side of the continuity and 
Euler equations (\ie $- 3H \rho$ and  $-H {\bf u}_p$ respectively) 
are due to the expansion of the universe and the factor $1/a$ is due to 
differentiation with respect to the comoving coordinates $\bf x$:
$\nabla \equiv {\partial / \partial x_i} \equiv a\cdot 
{\partial / \partial r_i}$.
The initial conditions are small density and smooth velocity 
perturbations imposed on a homogeneous density distribution. 

As long as the amplitude of the density perturbations remains small 
their evolution can be analyzed in the linear approximation 
obtained by the linearization of the above equations. 
The exact solution of the linearized system has one growing mode
which is the major object of our analysis and two decaying modes 
that can be ignored. The velocity in the growing mode is a potential vector field
which is proportional to the gradient of the linear gravitational potential: 
${\bf v}_{\rm lin} \propto -\nabla \phi_{\rm lin}$. 
In   the linear regime the spatial structure of the perturbations 
(in the comoving coordinates) remains unchanged
and its amplitude is proportional to the growing solution $\delta \propto D(t)$ 
also determined by the cosmological model \cite{car-etal-92}. 
For example, $D(t) \propto a(t) \propto t^{2/3}$ in the Einstein - de Sitter model,
in general $D(t)$ is a monotonically growing function of time. 

Equations \ref{eq:cont}--\ref{eq:poisson} describing the evolution of inhomogeneities in the universe
become more convenient for further analysis after the following transformation of variables are made
\cite{sh-94}
\be
\rho = a^{-3} \eta, ~~~ {\bf u}_p = a \dot{D}\, {\bf v}, ~~~
 \phi =\left( {3 \over 2} \Omega_0 \dot{a}^2 D\right) \varphi.
 \label{eq:general-variables}
\ee
In addition, we shall use $D$ instead of time, $t$. The resulting set of equations becomes
\ba
{\partial \eta \over \partial D} + {\partial (\eta v_i) \over \partial x_i} &=&0, \nonumber \\ 
{\partial v_i \over \partial D} + v_k {\partial v_i \over \partial x_k} &=& 
-{3\over 2} {\Omega_0 \over D f^2}\left( {\partial \varphi \over \partial x_i} + v_i \right), \label{eq:basic-D}\\
{\partial^2 \varphi \over \partial x_i^2} &=& {\delta \over D}. \nonumber
\ea
Here $\Omega_0 = \bar{\rho}_{m}/ \rho_c$ at present time (it is the ratio of the total mean mass 
density to the critical value $\rho_c=3H_0^2/8\pi G$), $f = d \ln D / d \ln a$ and
$\delta = (\eta - \bar{\eta})/\bar{\eta}$ where $\bar{\eta}=\Omega_0\rho_c = const$; 
we also assume the summation over repeated indices.

The second  equation in set (\ref{eq:basic-D}) becomes even simpler after introducing the total
derivatives $d/dD \equiv \partial / \partial D + v_k \partial /\partial x_k$
\be
{d v_i \over dD} = -{3\over 2} {\Omega_0 \over D f^2}\left( {\partial \varphi \over \partial x_i} + v_i \right).
\label{eq:lagr-accel}
\ee
One can easily check that in the linear regime (\ie while $<\delta^2>^{1/2} \ll 1$) the only growing 
solution to the set of equations (\ref{eq:basic-D}) is given by 
\ba
\delta({\bf q},D) &=& D {\partial^2 \Phi_0({\bf q}) \over \partial q_i^2}, \nonumber \\
{\bf v}({\bf q},D) &=& {\bf v}_0({\bf q}) = -\nabla_q \Phi_0({\bf q}),  \label{eq:lin-sol}\\
\varphi({\bf q},D) &=& \Phi_0({\bf q}), \nonumber
\ea
where ${\bf q}$ is the Lagrangian coordinates of a fluid element. The solution (\ref{eq:lin-sol})
gives the density contrast, velocity and potential in terms of the Lagrangian coordinates.
In order to find these fields in the Eulerian space one needs to solve the equation of trajectories
\begin{equation}
{\bf x}({\bf q},D)={\bf q}+D\cdot {\bf v}_0({\bf q})
\label{eq:Z-appr}
\end{equation}
for ${\bf q}={\bf q}({\bf x}, D)$ and substitute it in solution (\ref{eq:lin-sol}). 
The field $\Phi_0({\bf q})$
is determined by the perturbation of density in the linear regime which is assumed to be a realization
of a Gaussian random field. The power spectrum of the field that solely defines all its statistical
properties is derived from observations of the 
fluctuations of temperature in the Cosmic Microwave Background radiation (see \eg \cite{muk-05});
it is shown in Fig.\ref{fig:power-sp}. 

Although velocity $v_i$ remains constant for each particle in the linear regime the physical velocity
${\bf u_p}$ varies with time (Eq.\ref{eq:general-variables}). In the growing solution (Eq.\ref{eq:lin-sol}) gravitational force proportional to $-\partial \varphi / \partial x_i$, exactly balances the drag force
due to the uniform expansion of the universe $H{\bf u}_p \propto {\bf v}$ and thus the right hand side 
of Eq.\ref{eq:lagr-accel} vanishes to linear order.

\section{The Zel'dovich Approximation}
Zel'dovich (1970) derived the Lagrangian form of the linear theory and extrapolated it to the beginning  of the nonlinear regime \cite{z-70} (see also \cite{sh-z-89}).  This extrapolation
assumes that the initial perturbations are smooth in the sense that the initial power spectrum
must decrease faster than $k^{-3}$ at  $k \rightarrow 0$. In realistic cosmological
models \eg in the $\Lambda$CDM model the initial power spectrum $P(k) \propto \ln(k)k^{-3}$ 
up to very large $k$ corresponding to cosmologically irrelevant scales and thus the Zel'dovich approximation must be properly modified in order to be useful on the scale of the clusters and 
superclusers of galaxies. This will be discussed later.

\begin{figure*}[h]
    \epsfxsize = 14.2 cm
    \hskip 2 cm
    \epsfbox{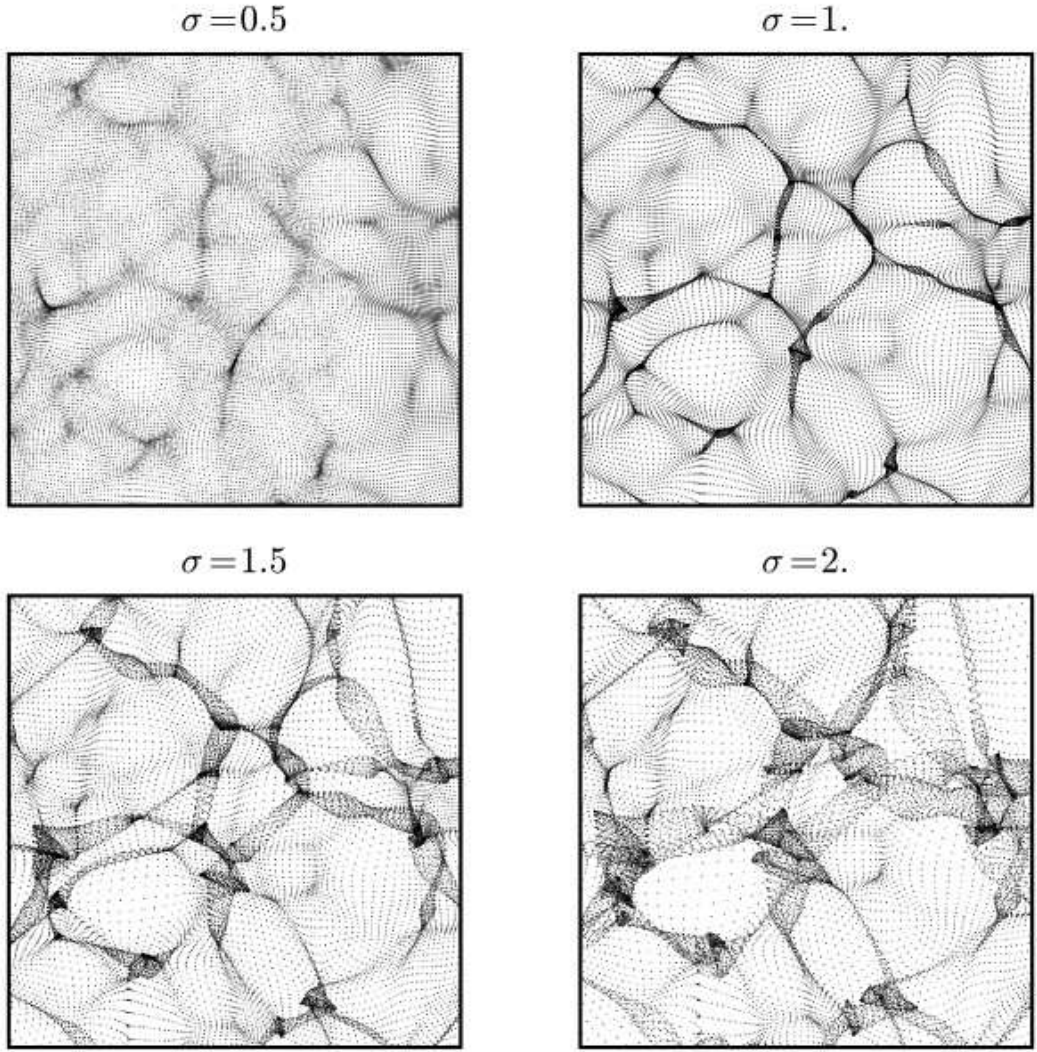}
    \caption{\small The distributions of particles at four stages of evolution described by the  
    Zel'dovich approximation in two dimensions. The panels are marked by $\sigma \equiv <\delta_{\rm lin}^2>^{1/2}$,
    the variance of the density contrast evaluated to the linear order, so $\sigma \propto D(t)$. } 
    \label{fig:z-4}
\end{figure*}

The Zel'dovich approximation can be conveniently formulated as a mapping from
Lagrangian space $L\{ q\} $ to Eulerian space $E\{ x\} $ defined by Eq.\ref{eq:Z-appr}.
 A two-dimensional illustration of the mapping is shown at four stages of the evolution
 in Fig.\ref{fig:z-4}. The stages are marked by $\sigma = <\delta_{\rm lin}^2>^{1/2}$ where 
 $\delta \equiv [\rho({\bf x}, t) -\bar{\rho}(t)] /\bar{\rho}(t) =  [\eta({\bf x},D(t)) -\bar{\eta}] /\bar{\eta}$
 is the density contrast. By convention, the parameter $\sigma$ is computed in the linear theory
 and therefore $\sigma \propto D(t)$.

\begin{figure}[h]
    \epsfxsize = 11 cm
    \hskip 2 cm
    \epsfbox{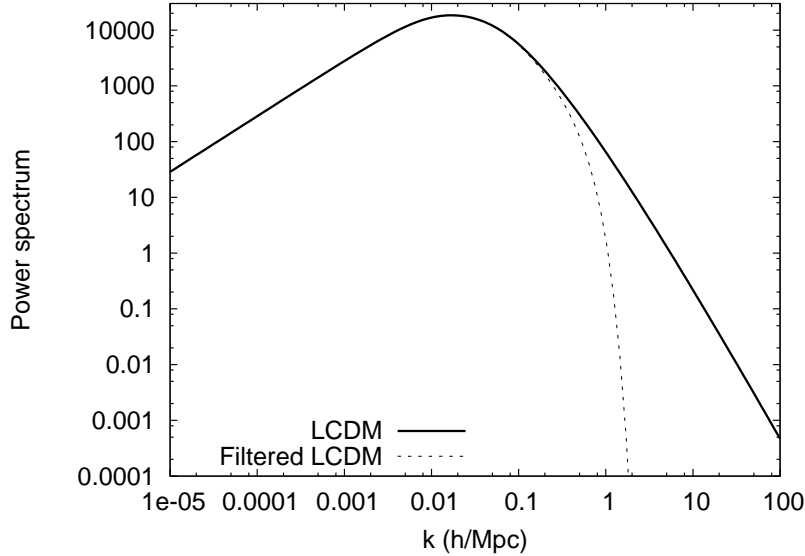}
    \caption{\small The initial power spectrum in the $\Lambda$CDM cosmology is shown
    by the solid line, the spectrum filtered at the current scale of nonlinearity is shown by the
    dotted line.} 
    \label{fig:power-sp}
\end{figure}

Utilizing conservation of mass $\eta d^3x = \bar{\eta}d^3q$ Zel'dovich has derived the
density as a function of $D(t)$ and the Lagrangian  coordinate ${\bf q}$
\begin{equation}
\eta({\bf q},D) ={\bar{\eta} \over {[1-D(t) \lambda_1({\bf q})]
[1-D(t) \lambda_2({\bf q})]
[1-D(t) \lambda_3({\bf q})]}}, \label{eq:rho}
\end{equation}
where $\lambda_1({\bf q})$, $\lambda_2({\bf q})$ and $\lambda_3({\bf q})$
are the eigenvalues of the deformation tensor $d_{ij}=\partial^2 \Phi_0/
\partial q_i \partial q_j$. 
In cosmology the initial condition is usually 
characterized by the spectrum $P_{\delta}(k)$ (Fig.\ref{fig:power-sp});
it is obviously related to the power spectrum of the potential as $P_{\delta} =  k^4 P_{\Phi_0}$.
 Similarly to the linear stage  one can find the density distribution in the Eulerian space
 by solving Eq.\ref{eq:Z-appr} for ${\bf q}$ and substituting it in Eq.\ref{eq:rho}. For realistic initial
 conditions it requires numerical calculations.

It follows from Eq.\ref{eq:rho} that the first objects
are formed around the peaks of the largest eigenvalue 
(we assume that 
$\lambda_{\rm i}$ are ordered at every ${\bf q}$: $\lambda_1 \ge \lambda_2$ and 
$\lambda_2 \ge \lambda_3$) and have
very oblate shapes because generic peaks of $\lambda_1$ always have two other eigen values
different from $\lambda_1$.
These objects are known in cosmology as ``Zel'dovich's pancakes''. 
This prediction of the Zel'dovich approximation differs  from the extrapolation of the Eulerian  
linear theory that predicts the formation of the first objects from the peaks of
$\delta = \lambda_1+\lambda_2+\lambda_3$.  In practice the difference in most cases
is not large due to obvious correlation between $\delta$ and $\lambda_1$ that causes the
peaks of $\delta$ be spatially close to the peaks of $\lambda_1$.
Three-dimensional gravitational N-body 
simulations are in a good agreement with the prediction of the Zel'dovich approximation 
\cite{sh-etal-95}.  Pancakes in the collisionless dark matter 
originate as the three-stream flow regions bounded by caustics, the surfaces
of formally infinite density. The shape and other characteristics of the
pancakes as well as of other generic types of structures (Fig.\ref{fig:z-4}) 
are determined  by catastrophe theory \cite{arn-sh-z-82}.

\begin{figure}[t]
    \epsfxsize = 18 cm
    \epsfbox{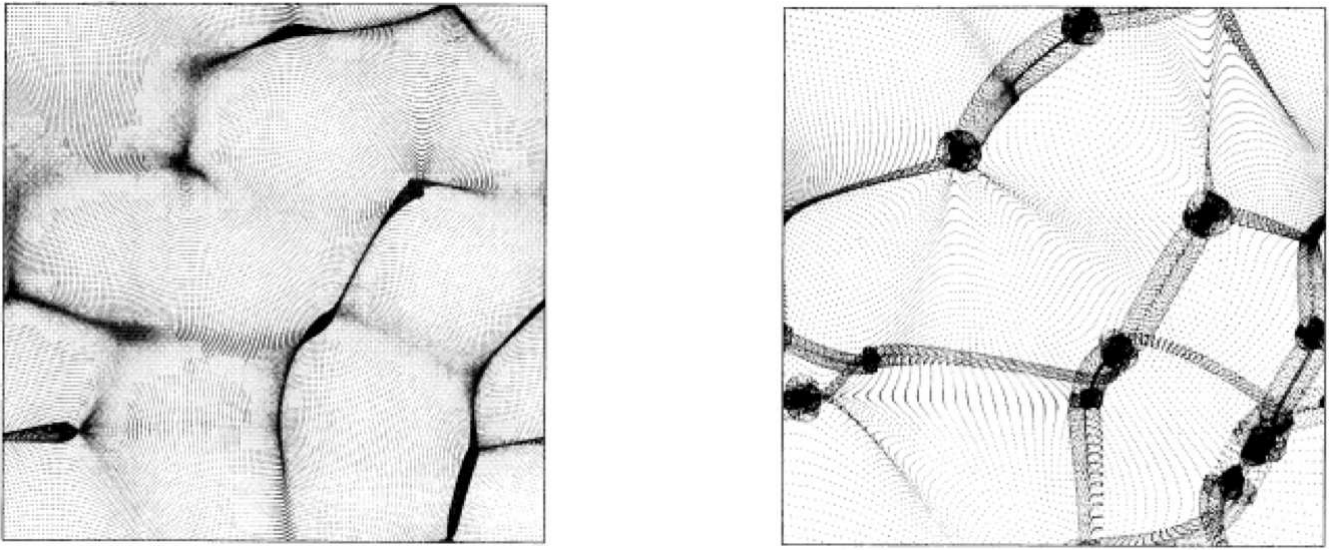}
    \caption{\small The distribution of particles at two stages $\sigma=1$ on the left and 
    $\sigma=2$ on the right obtained in a two-dimensional N-body simulation. 
    Adopted from \cite{sah-etal-94}.} 
    \label{fig:n-body-2d}
\end{figure}

The Zel'dovich approximation proved to be very good until orbit
crossing when caustics form and the multi-stream flows occur \ie up to the stage 
corresponding to $\sigma \approx 1$ (see e.g. \cite{sh-z-89} and references therein). 
At $\sigma > 1$ Eq.\ref{eq:Z-appr} predicts  broadening of the multi-stream flow regions 
considerably faster
than it happens  in N-body simulations \cite{dor-etal-80,kly-sh-83,nus-dek-90}
as can be seen in Fig. \ref{fig:n-body-2d}. Compare two panels on the right in Fig.\ref{fig:z-4}
with two panels  of  Fig. \ref{fig:n-body-2d}. Despite the difference in the realizations of the
initial conditions
one can see that the correct gravitational gravitational force in the N-body simulations results
in much thinner pancakes than predicted by the Zel'dovich approximation.
This discrepancy  motivated the development of the adhesion model that will be discussed  in the
following section. 
Here we briefly describe a simple modification that allows to utilize the Zel'dovich approximation
in the case when the small scale power cannot be ignored \cite{col-etal-93,mel-pel-sh-94}.

N-body simulations both in 2D and 3D have shown that the largest structures depend very little on
small-scale perturbations \cite{bea-etal-91,lit-wei-par-91,mel-sh-93}. The most important factor
that affect the shapes and general appearance of the large-scale structure are the initial waves
just reached nonlinear regime, \ie on scales where rms fluctuations are just becoming nonlinear
\cite{lit-wei-par-91}.
This observation allows to use an auxiliary model that yields the large-scale structures similar to
the original model but has no small-scale structures \cite{col-etal-93,mel-pel-sh-94}. 
The scale that separates large-scale structures
from small-scale structure is the scale of nonlinearity. 

For instance, the initial power spectrum
in the currently popular  $\Lambda$CDM cosmological model  is shown in Fig.\ref{fig:power-sp}
by the solid line while the spectrum filtered at the present scale of nonlinearity is shown by the dotted line. If the realization of amplitudes and phases are same in both models then the auxiliary model
with filtered power spectrum would have a very similar large-scale structure to one in the model 
with unfiltered power spectrum. The dotted line in Fig.\ref{fig:power-sp} shows the spectrum filtered
with a Gaussian window; a simpler version of filtering when the power spectrum set to zero on
scales with $k > k_{\rm nl}$  gives a similar result. 
The auxiliary model with filtered power spectrum provides an option of using the Zel'dovich
approximation  and therefore better understanding of complex nonlinear processes that control
the structure formation. The price of this simplification consists in missing all structures on 
smaller scales that are abundant in the model with unfiltered initial spectrum. 
The adhesion model is an attempt to remedy this drawback of the Zel'dovich approximation.

Finally, it is worth noting that both the Zel'dovich approximation (Fig.\ref{fig:z-4}, top right panel) 
and N-body simulations (Fig.\ref{fig:n-body-2d}) suggest that the universe can be viewed
as an irregular  tessellation where 'tiles' are represented by voids, the regions of low mass
and/or galaxy densities, where the flow consists of only one stream.
\section{The Adhesion Model}
The general idea of the adhesion model is very simple. We wish to use
the Zel'dovich approximation everywhere except the regions of multi-stream flows.
By adding a diffusion term to the Euler equation one can avoid the orbit crossing and
therefore suppress the  formation of the multi-stream flow regions. The advantage of
this modification is the suppression of too fast growth of the thickness of pancakes
observed in the Zel'dovich approximation but it is achieved at the coast  of noticeable
change of the dynamics inside multistream flow regions. A small coefficient of viscosity
guarantees agreement with the Zel'dovich approximation before the orbit crossing.

Since the motion is 
potential one can introduce the velocity potential ${\bf v}= -\nabla\Phi$. Then
assuming that the gravitational potential is approximately equal to the
velocity potential $\varphi \approx \Phi$  and also adding a viscosity term 
$\nu \nabla^2{\bf v}$  to the right hand side of the second
equation in set \ref{eq:basic-D} one obtains the equation of  nonlinear diffusion
\cite{gur-sai-sh-85,gur-sai-sh-89}
\begin{equation}
{\partial{\bf v}\over \partial D} + ({\bf v} \cdot \nabla){\bf v}
=\nu \nabla^2{\bf v}. \label{eq:burgers}
\end{equation}
Generally speaking the viscosity term needs not to be 
in the form of Eq.\ref{eq:burgers} but choosing this particular form 
one obtains Burgers' equation that has an exact analytic solution
\cite{bur-74}. For the potential motion ${\bf v}= -\nabla\Phi$ Eq.\ref{eq:burgers}
can be solved by performing the Hopf-Cole substitution $\Phi({\bf x},D) =
-2\nu \log U({\bf x},D)$. As a result Eq.\ref{eq:burgers} adopts the form 
of the linear diffusion equation
\begin{equation}
{\partial U \over \partial D} =\nu \nabla^2 U. \label{eq:diff}
\end{equation}
Solving Eq.\ref{eq:diff} for the velocity one obtains
\begin{equation}
{\bf v}({\bf x},D)
={{\int d^3q~\left[ ({\bf x}-{\bf q}) / D \right]~
\exp\left[S({\bf x},D;{\bf q})/2\nu \right]}
\over {\int d^3q~\exp[S({\bf x},D;{\bf q})/2\nu]}}, \label{eq:burg-sol}
\end{equation}
where the ``action''
\begin{equation}
S({\bf x},D;{\bf q})=\Phi_0({\bf q}) -{({\bf x}-{\bf q})^2 \over 2D}.
\end{equation}
\begin{figure}[h]
    \epsfxsize = 10 cm
    \hskip 4 cm
    \epsfbox{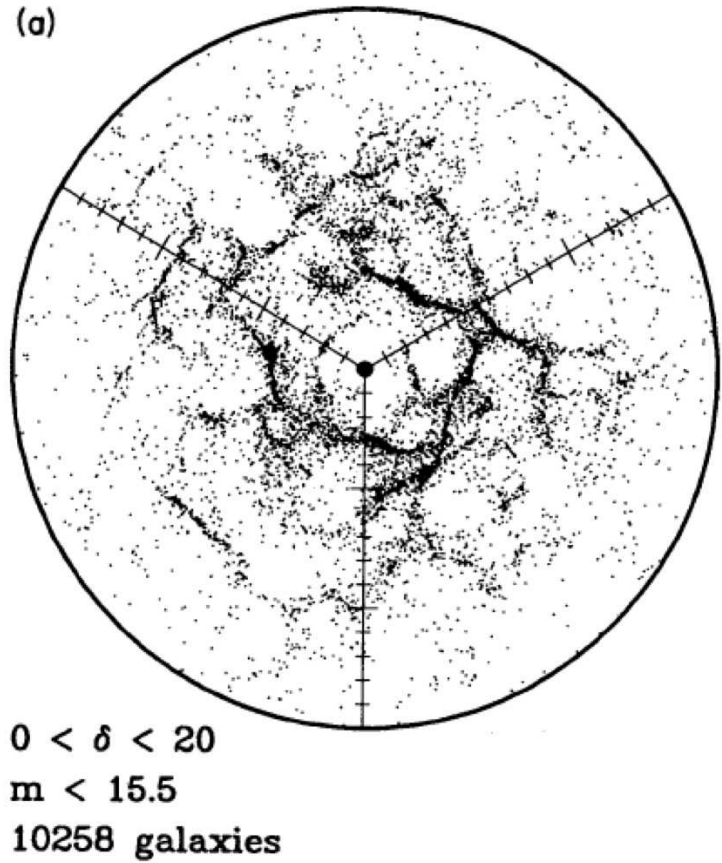}
    \caption{\small  Simulation of galaxy distribution in redshift space using the adhesion approximation with small but finite viscosity.
     Adopted from \cite{wei-gun-90}.}
   \label{fig:wei-gun-90}
\end{figure}

In cosmology the adhesion model has been used in two forms: one assumes
a small but finite value of the viscosity parameter $\nu$ and the
other assumes it is infinitesimal.
For small but finite $\nu$ the trajectory of a particle can 
be determined by solving the integral
equation and the resulting density can be determined from the continuity equation
 \cite{wei-gun-89,wei-gun-90,nus-dek-90,mel-sh-wei-94}. An example of the structure
 obtained by this method is shown in Fig.\ref{fig:wei-gun-90} where the authors \cite{wei-gun-90} 
 plotted the simulated galaxy distribution in a slice through the simulation box similar to 
 the plots obtained from galaxy redshift catalogs as in Fig.\ref{fig:GWs}. An interesting 
 attempt to explain the magnitude of the coefficient of viscosity from dynamics in multistream
 flow regions was made in \cite{buc-dom-05}.
\begin{figure}[h]
    \epsfxsize = 10 cm
    \hskip 4. cm
    \epsfbox{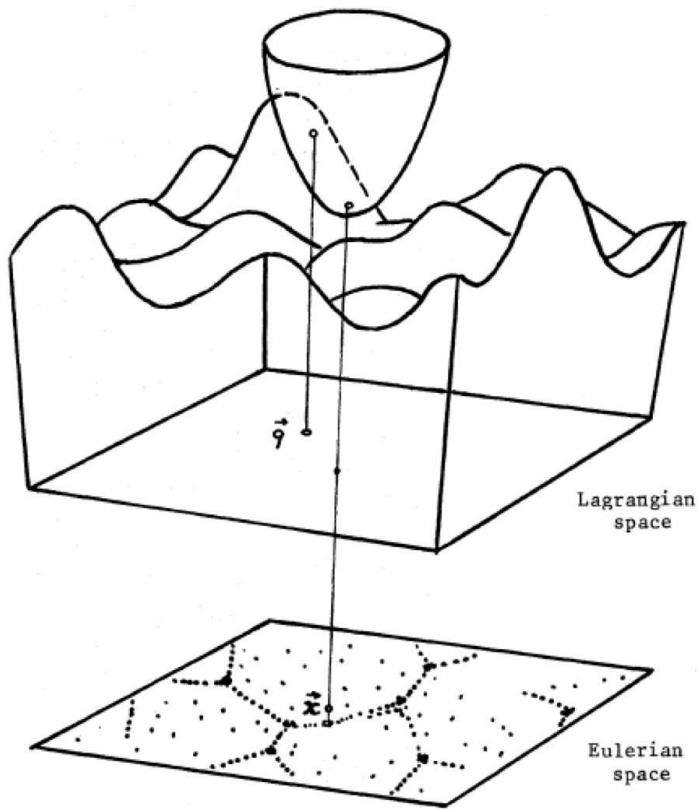}
    \caption{\small The illustration of the geometrical technique that finds the Eulerian position of a
     particle using the initial potential $\Phi_0({\bf q})$. For a given value of $D$ the paraboloid
     (Eq.\ref{eq:paraboloid}) descends on the surface of the initial potential $\Phi_0$ until it is
     tangent to the surface at point ${\bf q}$ by adjusting constant $P_0$. The projection of the 
     apex of the paraboloid denotes the Eulerian position of the particle.  
     Adopted from \cite{kof-pog-sh-90}.} 
    \label{fig:paraboloid-2d}
\end{figure}

We will discuss a different approach that assumes the infinitesimal value of the viscosity parameter 
$\nu \rightarrow 0$ in more detail because it has  direct connection to tessellation and tiling.
In this case the integrals in Eq.\ref{eq:burg-sol} can be evaluated using the method of 
steepest descents \cite{gur-sai-sh-85,gur-sai-sh-89,kof-pog-sh-90,kof-etal-92}. 
The result has a remarkable geometrical 
interpretation illustrated by Fig.\ref{fig:paraboloid-2d} in two-dimensional case. 
The initial Lagrangian coordinates 
${\bf q}_p$ and the current Eulerian coordinates ${\bf x}_p$ of a particle at a chosen stage
denoted by $D$,  can be related simply by descending a paraboloid 
\begin{equation}
P({\bf x},D;{\bf q}) ={({\bf x}-{\bf q})^2 \over D} + P_0,
\label{eq:paraboloid}
\end{equation}
with apex at ${\bf x} ={\bf x}_p$ and adjusting correspondently the constant $P_0$
until it is in contact with the surface of the initial potential $\Phi_0({\bf q})$ at some point. 
The coordinates of the contact point are the Lagrangian coordinates of the particle ${\bf q}_p$.
This procedure must satisfy an additional constraint:  the paraboloid must not cross $\Phi_0$
at any point. Thus, the rule is: the tangential contacts of the paraboloid with the potential
are welcome  but crossings are strictly forbidden. 
At the linear stage $D$ is small and the curvature of the paraboloid is greater
than the typical curvature  of $\Phi_0$, thus this condition can be easily fulfilled.
The Zel'dovich approximation is universally valid at this stage  and
 the velocity filed is given by the following equation
\begin{equation}
{\bf v}({\bf x},D)={{\bf x}-{\bf q}({\bf x},D) \over D}, \label{eq:burg-vel}
\end{equation}
where ${\bf q}({\bf x},D)$ is the coordinate of the absolute maximum of
the action $S({\bf x}, D; {\bf q})$ at given ${\bf x}$ and $D$. 
\begin{figure}[t]
    \epsfxsize = 4.5 truein
    \hskip 1.5 truein
    \epsfbox{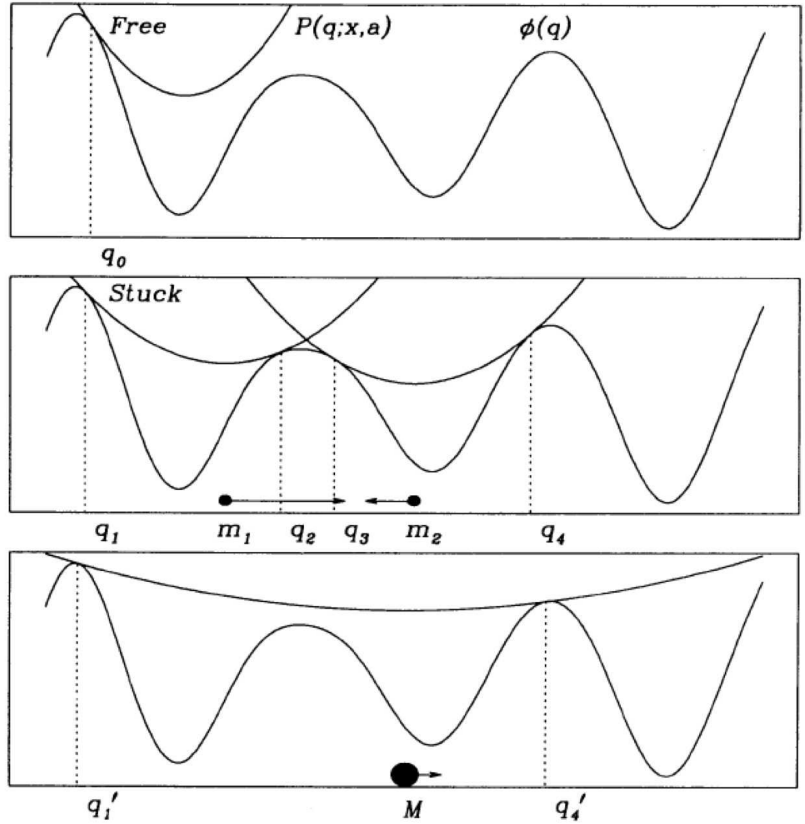}
    \caption{\small An illustration of the geometrical prescription of descending 
    a parabola onto the initial velocity potential in order to find the Eulerian positions
     of particles and knots in one-dimenaional case. The particle having
     Lagrangian coordinate $q_0$ in the uppermost panel has
      the Eulerian coordinate of the paraboloid apex. 
       In the middle panel corresponding to a later stage 
       the whole region between two points where the parabola is tangent to the initial potential
        is stuck into the knot $m_1$ moving with the velocity shown
        by the arrow, its mass is $m_1 =\bar{\eta}(q_2-q_1)$. 
        The knot $m_2$ is determined by the second paraboloid in the
        middle panel.
       The lower panel shows the knot $M$ formed as a result of merging $m_1$,
         $m_2$ and the mass between them. 
         Adopted from \cite{sah-etal-94}.}
    \label{fig:parabolas-1d}
\end{figure}

As time passes the curvature of the paraboloid decreases, so it becomes wider. 
As a result the points ${\bf q}$ that are not accessible without violation of the crossing rule 
as can be seen in one-dimensional illustration in  Fig.\ref{fig:parabolas-1d}. 
Both parabolas shown in the middle panel
cannot descend further without crossing the initial potential $\Phi_0$.
What has happened to the corresponding particles?
The particles between two contact points \ie between $q_1$ and $q_2$ for one
parabola and between $q_3$ and $q_4$ for the other 
have been stuck into point masses $m_1$ and $m_2$ respectively 
due to orbit crossing.
Their coordinates are the coordinates of the apices of the corresponding parabolas
and the masses are $m_1=\bar{\eta} (q_2 -q_1)$ and
$m_2=\bar{\eta} (q_4 -q_3)$. At later times the discrete  masses 
formed earlier can grow by merging  as well as 
by accreting the continuous medium between them. This is  illustrated by the
bottom panel. It is not hard to see that the transition from
the state shown in the middle panel to one shown in the bottom panel requires the state 
when one parabola has three contacts with the potential simultaneously. 
At the next instant of time it becomes
a little wider and looses the contact with the central peak.

In two dimensions the paraboloid can be tangent to the initial potential $\Phi_0$ simultaneously
in one, two or three points. The first type corresponds to the particles that form continuous 
medium. The points of the second type have been stuck into lines
shown by thick points in  Fig.\ref{fig:paraboloid-2d}. These lines, the edges of tiles, 
are formed by the set of the apices of the paraboloids that are in contact with the initial potential 
at two points simultaneously. Finally, when the paraboloid is in contact with the surface of $\Phi_0$
at three point simultaneously its apex marks the knot (vertex) where the lines (edges) meet. 
The edges and vertices form
the boundaries of irregular tiles of tessellation in two dimensions. 
In the course of the evolution some tiles expand while the other collapse and cease to
exist. This goes hand in hand with  merging of the knots. At the instant of a merger 
the paraboloid can be in contact with the initial potential at more than three points simultaneously
\cite{arn-bar-bog-91}.
\begin{figure}[t]
    \epsfxsize = 17 cm
    \hskip 1.5 cm
    \epsfbox{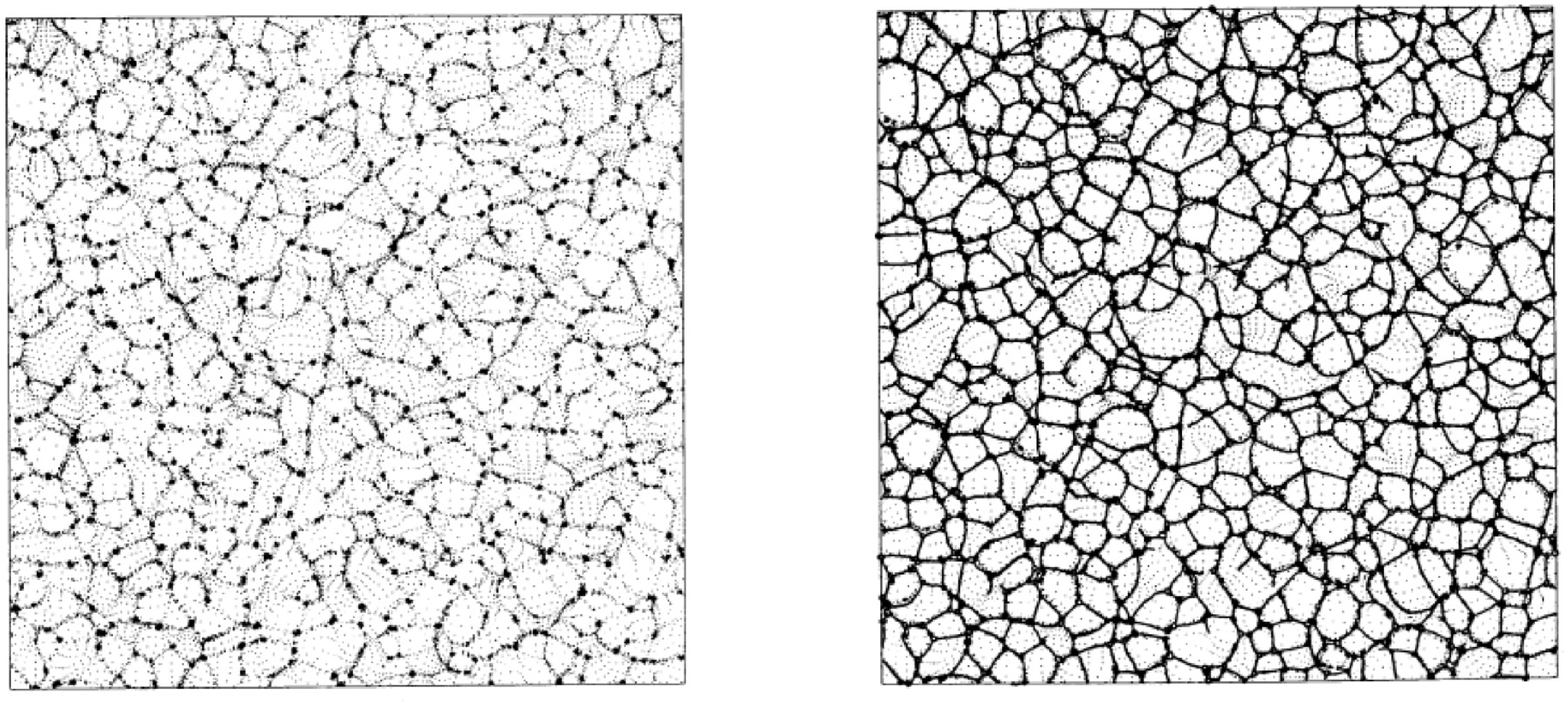}
    \caption{\small  The left hand side panel shows the distribution of particles obtained in
    two-dimensional N-body simulation. The panel on the right shows the tessellation
    computed in the adhesion model from the same initial conditions. The tessellation is
    superimposed on the particle distribution. Adopted from \cite{kof-etal-92}.}
   \label{fig:nb-vs-aa}
\end{figure}

Figure \ref{fig:nb-vs-aa} demonstrates the agreement between the tessellation computed
in the adhesion model and the N-body simulations in two-dimensional space.
\begin{figure}[h]
    \epsfxsize = 17 cm
    \hskip 1.5 cm
    \epsfbox{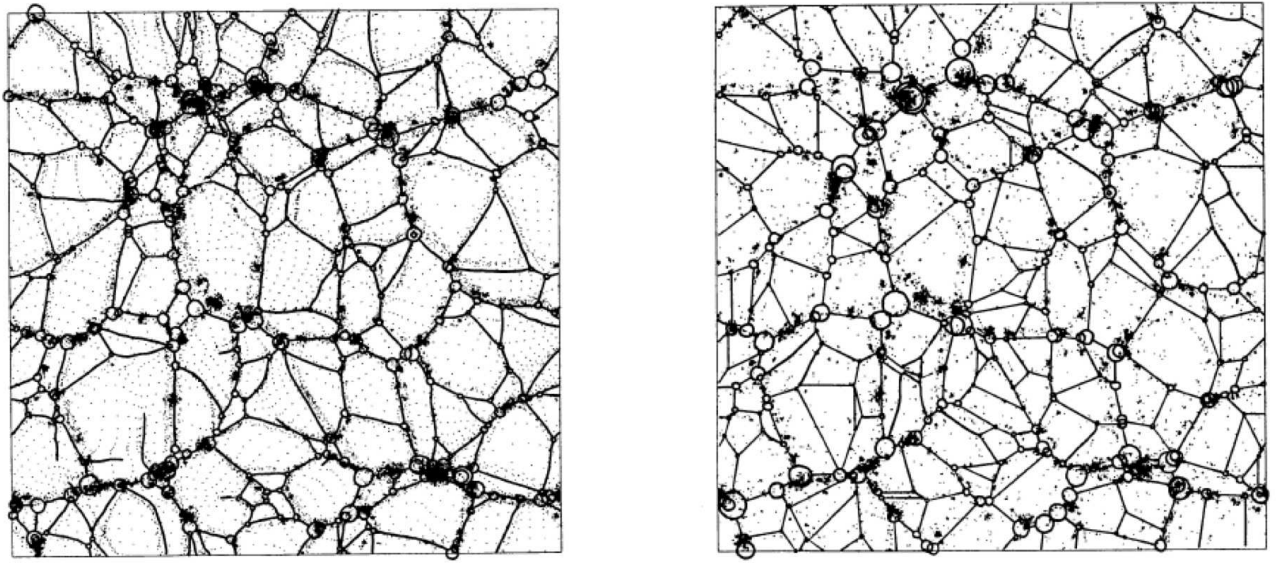}
    \caption{\small Two panels show the tessellation of the plane by the adhesion model
    superimposed on the particle distribution obtained in 2D N-body simulation.  The initial
    potential $\Phi_0$  in both panels have the same large-scale part but on the left hand side 
    it does not have the small-scale part that is present in the potential on the right hand side.
     Adopted from \cite{kof-etal-92}.}
   \label{fig:nd-vs-aa-diff-cutoffs}
\end{figure}
Figure 
\ref{fig:nd-vs-aa-diff-cutoffs} illustrates the effect of small-scale power. The initial potentials
in both panels were generated with the same amplitudes and phases but the amplitudes
 for the short waves $k > k_{cr}$ were set to zeros in the case shown in the left hand side panel.
 Therefore, the initial potential in the left hand side panel is a smooth version of the potential
 in the right hand side panel.
The both models evolved to the same stage. One can see that while the overall appearance of 
the tessellation is similar in both panels the there more small-scale details in the panel on the right.

With a little imagination and help from  the geometric model outlined above one can easily 
visualize the main features of the tessellation in three-dimensional space.
The initial potential $\Phi_0$ forms a three-dimensional hypersurface in four-dimensional
space. A three-dimensional paraboloid (Eq.\ref{eq:paraboloid}) descends on the 
thre-dimensional hypersurface of the initial potential similar to one- and two-dimensional cases. 
The rules are the same: it can touch the initial potential but must not cross it. 
Four generic situations are possible at a generic
instant of time: it can have only one, two, three or four contact points simultaneously.
If there is only one contact point then the corresponding particle is in free motion; 
the coordinates of the  of contact point are the Lagrangian coordinates of the particle 
and the apex denores its Eulerian coordinates.
In the remaining three cases the apex define a particle stuck on a surface, on an edge or at a vertex
of the tessellation, if the number of contact points is two, three or four respectively. Again
similarly to one- and two-dimensional cases at the critical instants of time when a
tree-dimensional tile collapses and ceases to exist or two or more vertices  merge the number 
of contact points can be greater than four \cite{arn-bar-bog-91}.

The set of surfaces, filaments and knots define the tessellation of three-dimensioal
space: voluminous low density tiles (voids) are separated 
by the faces (pancakes), the edges  (filaments) are at the intersections of the faces,
and the vertices (clusters of galaxies, halos in N-body simulations)  are at the intersections 
of the edges. This tessellation can be viewed as the skeleton of the real structure.

The both variants of the adhesion model (with small but finite and infinitesimal viscosity) 
have been tested against the  gravitational N-body simulations in two and 
three dimensions. Both the N-body simulation
and the adhesion model used the identical initial conditions and were compared
at several stages of the evolution. 
The geometrical version of the adhesion model was tested against the 
two-dimensional N-body
simulations with the initial power law spectra $P_\delta(k) \propto k^n$
with spectral indices $n=2,0$ and $-2$ and various cutoffs 
\cite{kof-etal-92}.
The version of the adhesion model utilizing a finite viscosity
parameter $\nu$ has been
quantitatively compared to fully nonlinear, numerical three-dimensional
gravitational N-body simulations in \cite{mel-sh-wei-94}. The results 
generally show a good agreement however they improve with the growth
of the relative power on large scales.  This is the case in the $\Lambda$CDM model  
 as one can see from the initial power spectrum shown in Fig.\ref{fig:power-sp}. 
\section{Summary}
\begin{figure}[h]
    \epsfxsize = 10 cm
    \hskip 4 cm
    \epsfbox{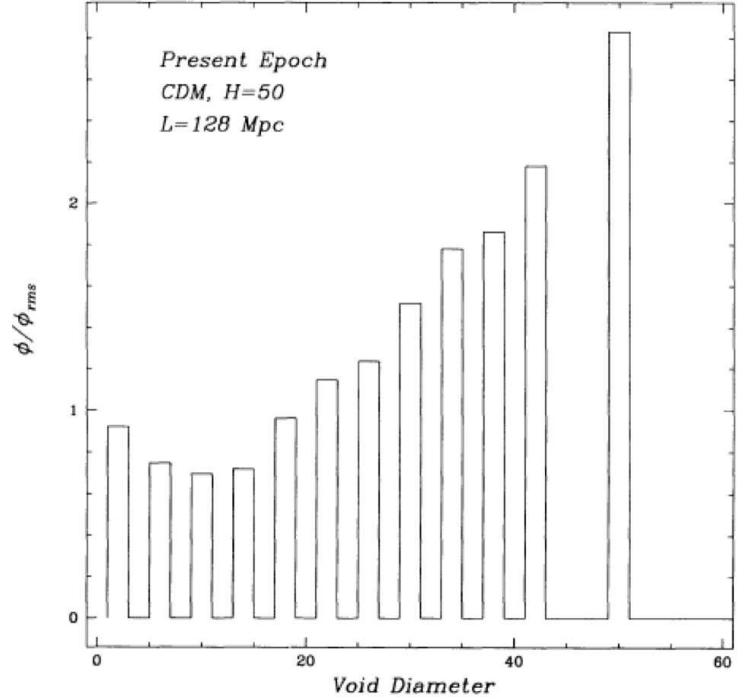}
    \caption{\small Normalized initial gravitational potential at the centers of voids is plotted
    against the void diameter.  Adopted from \cite{sah-etal-94}.}
   \label{fig:voids}
\end{figure}
The adhesion model based on three-dimensional Burgers' equation (Eq.\ref{eq:burgers})
of the nonlinear diffusion has been found to provide a good explanation
of the major large-scale features of the galaxy distribution in  the universe. 
Its particular version assuming the infinitesimal coefficient of viscosity $\nu \rightarrow 0$
provides a natural geometrical construction that can be characterized 
as "quasi-Voronoi tessellation" \cite{mol-etal-97}.
Comparison with gravitational two- and three-dimensional N-body
simulations has shown a good agreement including advanced nonlinear stage.
The adhesion model provides a natural qualitative explanation of
the origin of the large-scale
coherent structures such as superpancakes and superfilaments, as a result
of coherent motion of clumps due to large-scale inhomogeneities in the
initial gravitational potential \cite{kof-sh-88}.  The initial gravitational potential
also affects the sizes of voids as shown in Fig.\ref{fig:voids}: the higher the peak
of the initial potential the greater size of the voids formed around it.
The formation and evolution of large-scale structure is described by the
adhesion model as a two stage process \cite{sah-etal-94}. During the first short
stage matter falls predominantly onto pancakes and then moves along them towards filaments 
and then along filaments to the  knots.  At the end of the first
stage the formation of the tessellation is completed 
and  the bulk of the mass in the universe is located in 
knots (vertices) and filaments (edges)
however a relatively small fraction of mass remains in pancakes (faces)
and voids (three-dimensional tiles). The second stage is characterized by the
deformation of the tiles and edges  due to the dynamics caused by gravity . 
At this stage knots
merge into more massive knots and small voids shrink and cease to exist giving space 
to growing  large voids. Eventually almost all the mass concentrates in knots.
Depending on the initial spectrum the knots themselves may move coherently in such a
manner that they concentrate to superpancakes and superfilaments. 
This situation is  likely to occur in the \lcdm model at the present time 
due to the shape of the initial power spectrum (Fig.\ref{fig:power-sp}).
The superpancakes and  superfilaments can be identified by applying the
adhesion model to {\it smoothed} initial potential \cite{kof-etal-92}.

{\bf Acknowledgement} The author is grateful to Rien van de Weygaert for many useful discussions 
related to the subject of this review. 
\vspace{0.5in}

\vfill
\end{document}